\documentclass[conference]{IEEEtran}

 \usepackage[pdftex]{graphicx}

 \DeclareGraphicsExtensions{.pdf,.jpeg,.png}

\begin{document}

\title{A formal experiment to assess the efficacy of certification standards}

\author{\IEEEauthorblockN{Virginie Wiels}
\IEEEauthorblockA{ONERA-DTIM
2 avenue E. Belin, BP74025\\
31055 Toulouse cedex, France\\
Email: Virginie.Wiels@onera.fr}
}


%


\maketitle

\begin{abstract}
Proving the efficacy of certification standards?
\end{abstract}


\section{Introduction}
Having a formal background, the unplanned experiment I would like to propose to assess the efficacy of standards for safety critical software would be to formally prove the following theorem: "if a software conforms to a standard, then it is correct". My favorite application domain is aerospace, so I propose to place this experiment in the DO-178 context. I do not pretend this "experiment" is possible, but I hope that trying to formally express the problem will help in understanding some things about it (as it is usually the case when using formal methods in general).

\section{Certification goal and means}

In this section, we first describe the different  elements that are at stake in a certification context. We state the goal of certification, describe existing means with respect to that goal, in order to try to make the link between goal and means in the next section. 

\subsection{Goal of certification }
 The item that is under scrutiny is the embedded software, it is the one that is the target of certification (the one that should execute correctly). And the goal of certification is to ensure that this software indeed executes correctly, with respect to a set of system requirements. We forget here abour some aspects of certification, such as planning, configuration management, quality assurance, we essentially focus on verification

\subsection{DO-178: existing means}

DO-178 \cite{do178} does not prescribe a specific development process, but instead identifies important activities and design considerations throughout a development process and defines objectives for each of these. DO-178 distinguishes development processes from “integral” processes that are meant to ensure correctness, control, and confidence of the software life cycle processes and their outputs. The verification process is part of the integral processes along with configuration management and quality assurance. 
Four processes are identified as comprising the software development processes in DO-178:
\begin{itemize}
\item	The software requirements process develops High Level Requirements (HLR) from the outputs of the system process;
\item	The software design process develops Low Level Requirements (LLR) and Software Architecture from the HLR;
\item	The software coding process develops source code from the software architecture and the LLR;
\item	The software integration process loads executable object code into the target hardware for hardware/software integration. 
\end{itemize}
Each of these processes is a step towards the actual software product. 

The results of the four development processes must be verified. Detailed objectives are defined for each step of the development, with some objectives defined on the output of a development process itself and some on the compliance of this output to the input of the process that produced it. For example, LLR shall be accurate and consistent, compatible with the target computer, verifiable, conformed to requirements standards, and they shall ensure algorithm accuracy. Furthermore, LLR shall be compliant and traceable to HLR.

\section{What we would like to prove}

If we want to formally assess the efficacy of DO-178, it would ideally consist in showing that:

\begin{itemize}
\item the identified parts of the development processes satify the certification objectives,
\end{itemize}

implies that

\begin{itemize}
\item the embedded software is correct.
\end{itemize}

\subsection{Making the link}

If we try to make the link between targetted goal and existing means, we can identify two dimensions and four different levels. The two dimensions are the representation of the software and the definition of objectives for this software. The four levels are the following:
\begin{itemize}
\item the embedded software itself,
\item all the elements of processes that were developed to produce this software (requirements, design, etc),
\item the certification argument that includes part of the development processes elements and the justification of the certification objectives,
\item the certification standard that specifies the expected content of the certification argument.
\end{itemize}



\subsection{Focus}

In the following of the paper, we will forget about some aspects of certification. We will in particular ignore the assessment of  the certification argument by certification authorities,this assessment involves human issues that we consider out of scope of our formal experiment. We will focus on the fact that the certification standard proposes to evaluate the embedded software by looking at some properties of some elements of the process used to produce this software. 

\section{Certification as a form of abstraction}

We propose to formalise a certification standard as a form of abstraction. In formal verification, when a property cannot be verified on an artifact, a possible solution is to define an abstraction that is applied to the artifact and the property in order to obtain a model that is tractable using the verification technique. Abstract interpretation for example \cite{ai77} proposes to define an abstraction of the semantics of a program, abstraction that is defined with respect to the property to be verified (for example, if the property addresses the sign of the result of a multiplication, all the variables can be abstracted by their sign).

\begin{figure}[]
\centering
\includegraphics[width=3.5in]{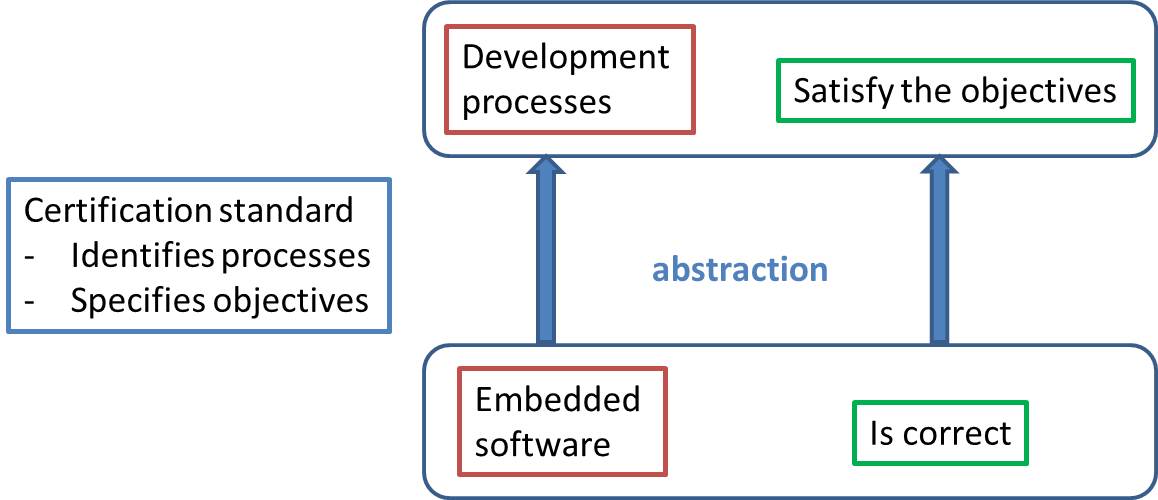}
\caption{Certification as an abstraction}
\label{fig_abs}
\end{figure}

Certification is the definition of an abstraction. As the correction of an embedded software cannot be verified, DO-178 defines an abstraction: the verification of correction is abstracted by the verification of objectives on HLR, LLR, software architecture, source code and executable object code. The standard is the definition of the abstraction itself, as represented on Figure \ref{fig_abs}.

\subsection{Charcateristics of an abstraction. }
Three characteristics are important for the definition of an abstraction:
\begin{itemize}
\item the abstraction is sound: it defines more behaviors than the actual behaviors of the program, so that when a property is verified on the abstract program,it is also true of the concrete one;
\item the abstraction is sufficiently abstract to compute properties efficiently;
\item it is sufficiently precise to provide good answers. When the abstraction is too wide, a certain number of potential errors are found in the abstract program that are not reachable by the concrete program, the so called {\em false alarms}.
\end{itemize}

Soundness is essential, the other two characteristics express a compromise between precision and tractability.



\subsection{Transposition to certification}
If we try to transpose the three charcateristics of abstraction to certification, it would require to assess the following questions:
\begin{itemize}
\item is the abstraction defined by certification sound? 
\item is it tractable in practice?
\item is it precise enough or does it imply a lot of false alarms?
\end{itemize}

The answer to the second and third questions are mostly positive. Even if the cost of DO-178 certification is high for the most critical certification levels, this certification is tractable in practice. Precision is not an issue in practice. There are no complaints on errors that would be detected at abstract level (by the certification process) and that would not be an effective  problem at the concrete level. It does not mean there is no such case, but that they are not an issue in practice (during the writing of DO-333 \cite{do333}, some debates took place on the necessity of the removing of dead code in case of formally proven code, this could be a potential example of false alarm). 

The answer to the first question is not so positive. There are cases where bugs were detected in software that had previously been certified. The abstraction defined by certification is thus not sound. However, it is difficult to know if these bugs were due to a justification of the certification objectives that was not sufficiently thorough (the activities defined to reach the objective were not sufficient); or to a lack in the definition of the certification objectives themselves. It may be that the abstraction defined by certification is sound but that the "computation" of this abstraction is not correct. In order to be able to distinguish between the two, it would be necessary to formalise this abstraction and to study its property (demonstrate soundness). 


\section{Conclusion}
If we consider certification as a form of abstraction, and try to determine if it is a "good" abstraction, we first can claim that it performs well on the usability aspect: certification looks like a good compromise between precision and tractability. However, a formalisation of this abstraction would be necessary to decide if it is a sound abstraction. And this formalisation would require in particular to formalise the notion of "correctness" of software (compliance to system requirements only?) which is important because we saw that an abstraction is in general dedicated to a given property.






\begin{thebibliography}{1}

\bibitem{do178}
RTCA/EUROCAE. DO-178C/ED-12C: Software Considerations in Airborne Systems and Equipment certification, 2011.

\bibitem{do333}
RTCA/EUROCAE. DO-333/ED-216: Formal Method Supplement to DO-178C/ED-12C, 2011.


\bibitem{ai77}
Patrick Cousot, Radhia Cousot: Abstract Interpretation: A Unified Lattice Model for Static Analysis of Programs by Construction or Approximation of Fixpoints. Conference Record of the Fourth ACM Symposium on Principles of Programming Languages, Los Angeles, California, USA, January 1977. ACM, 1977, pp. 238-252

\end{thebibliography}
%

\end{document}